\documentstyle[12pt,axodraw]{article}

\begin{document}

\parskip=0.3cm
\begin{titlepage}

\hfill \vbox{\hbox{DFPD 00/TH/48}\hbox{UNICAL-TH 00/7}
\hbox{October 2000}}

\vskip 0.2cm

\centerline{\bf FINITE SUM OF GLUON LADDERS}
\centerline{\bf AND HIGH ENERGY CROSS SECTIONS $~^\star$}

\vskip 0.3cm

\centerline{R.~Fiore$^{a\dagger}$, L.L.~Jenkovszky$^{b\ddagger}$,
E.A.~Kuraev$^{c\diamond}$,
A.I.~Lengyel$^{d\S}$, F.~Paccanoni$^{e\ast}$, A.~Papa$^{a\dagger}$}

\vskip 0.1cm

\centerline{$^{a}$ \sl  Dipartimento di Fisica, Universit\`a della Calabria 
\& INFN-Cosenza,}
\centerline{\sl I-87036 Arcavacata di Rende, Cosenza, Italy} 

\centerline{$^{b}$ \sl  Bogolyubov Inst. for Theor. Phys., Ac. of Sciences of 
Ukraine}
\centerline{\sl UA-03143 Kiev, Ukraine}

\centerline{$^{c}$ \sl JINR, RU-141980 Dubna, Russia}

\centerline{$^{d}$ \sl Institute of Electron Physics,}
\centerline{\sl Universitetska 21, UA-88000 Uzhgorod, Ukraine}

\centerline{$^{e}$ \sl Dipartimento di Fisica, Universit\`a di Padova
\& INFN-Padova,}
\centerline{\sl via F. Marzolo 8, I-35131 Padova, Italy}

\vskip 0.1cm

\begin{abstract}
A model for the Pomeron at $t=0$ is suggested. It is based on the idea of a
finite sum of ladder diagrams in QCD. Accordingly, the number of $s$-channel
gluon rungs and correspondingly the powers of logarithms in the forward
scattering amplitude depends on the phase space (energy) available, i.e. as
energy increases, progressively new prongs with additional gluon
rungs in the $s$-channel open. Explicit expressions for the total
cross section involving two and three rungs or, alternatively, three and
four prongs (with $\ln^2(s)$ and $\ln^3(s)$ as highest terms, respectively) 
are fitted to the proton-proton and proton-antiproton total cross section 
data in the
accelerator region. Both QCD calculation and fits to the data indicate 
fast convergence of the series. In the fit, two terms (a constant and a 
logarithmically rising one) almost saturate the whole series, the 
$\ln^2(s)$ term being small and the next one, $\ln^3(s)$, negligible.
Theoretical predictions for the photon-photon total cross section are also 
given.

\end{abstract}

\vskip 0.1cm

\hrule

\vskip 0.1cm

\noindent
$^{\star}${\it Work supported by the Ministero italiano dell'Universit\`a e 
della Ricerca Scientifica e Tecnologica and by the INTAS.}

$
\begin{array}{ll}
^{\dagger}\mbox{{\it e-mail address:}} &
   \mbox{FIORE,~PAPA@CS.INFN.IT} \\
^{\ddagger}\mbox{{\it e-mail address:}} &
\mbox{JENK@GLUK.ORG} \\
^{\diamond}\mbox{{\it e-mail address:}} &
\mbox{KURAEV@THSUN1.JINR.RU} \\
^{\S}\mbox{{\it e-mail address:}} &
   \mbox{SASHA@LEN.UZHGOROD.UA} \\
^{\ast}\mbox{{\it e-mail address:}} &
   \mbox{PACCANONI@PD.INFN.IT}
\end{array}
$

\end{titlepage}
\eject
\newpage

\textheight 210mm \topmargin 2mm \baselineskip=24pt

\section{Introduction}

It is widely accepted that the Pomeron in QCD corresponds to an infinite 
sum of gluon ladders with Reggeized gluons on the vertical lines (see Fig.~1), 
resulting~\cite{FKL,BL,L} in the so-called supercritical behavior 
$\sigma _t\sim s^{\alpha_P (0)}$, $\alpha_P (0) >1$, where $\alpha_P (0)$ is 
the intercept of the 
Pomeron trajectory. In that approach, the main contribution to the 
inelastic amplitude and to the absorptive part of the elastic amplitude 
in the forward direction arises from the multi-Regge kinematics in the 
limit $s \to\infty$ and leading logarithmic approximation. In the 
next-to-leading logarithmic approximation (NLLA), corrections require also
the contribution from the quasi-multi-Regge kinematics~\cite{FL98}.
Hence, the subenergies between neighboring 
$s$-channel gluons must be large enough to be in the Regge domain. At 
finite total energies, this implies that the amplitude is represented by 
a finite sum of $N$ terms~\cite{fiore}, where $N$ increases like $\ln s$, 
rather than by the solution of the BFKL integral equation~\cite{FKL,BL,L}.
The interest in the first few terms of the series is related to the 
fact that the energies reached by the present accelerators are not high 
enough to accommodate a large number of $s$-channel gluons that eventually 
hadronize and give rise to clusters of secondary particles.

The lowest order diagram is that of two-gluon exchange, 
first considered by Low and Nussinov~\cite{LN}. The next order, involving an 
$s$-channel gluon rung was studied e.g. in the papers~\cite{BL,McCoy}. 
The problem of calculating these diagrams is twofold. The first one is connected
with the nonperturbative contributions to the scattering amplitude in the
''soft'' region. It may be ignored by ''freezing'' the running coupling
constant at some fixed value of the momenta transferred and assuming that the
forward amplitude can be cast by a smooth interpolation to $t=0$. More
consistently, one introduces a nonperturbative model~\cite{Francesco} of the
gluon propagator valid also in the forward direction. The second problem is
more technical: as $s\to \infty $ the number of Feynman diagrams that contribute
to the leading order rapidly increases and, in each of them,
only the leading contribution is usually evaluated. At any order in the
coupling, subleading terms coming both from the neglected
diagrams and from the calculated ones are present. Although functionally the 
result is always the sum of increasing powers of logarithms, the numerical 
values of the coefficients entering the sum is lost unless all diagrams 
are calculated.

Conversely, one can expand the "supercritical" Pomeron $\sim s^{\alpha(0)}$
in powers of $\ln(s)$. Such and expansion is legitimate within the range of
active accelerators, i.e. near and below the TeV energy region, where fits
to total cross sections by a power or logarithms are known~\cite{Ezhela} to
be equivalent numerically. Moreover, forward scattering data (total cross
sections and the ratio of the real to the imaginary part of the forward
scattering amplitude) do not discriminate even between a single and
quadratic fit in $\ln(s)$ to the data.

Phenomenologically, more information on the nature of the series can be
gained if the $t$ dependence is also involved. The well-known (diffractive)
dip-bump structure of the differential cross section can be roughly imitated
by the Glauber series, although more refined studies within the dipole
Pomeron model (DP) (linear behavior in $\ln(s)$)~\cite{JShS} show that the
relevant series is not just the Glauber one. A generalization of the DP
model including higher powers of $\ln(s)$ was considered in~\cite{Pierre}. 
In a recent paper~\cite{bea} the Pomeron was considered as a finite 
series of ladder diagrams, including one gluon rung besides the 
Low-Nussinov "Born term" and resulting in a constant plus logarithmic 
term in the total cross section. With a sub-leading Regge term added, 
good fits to $pp$ and $p\bar p$ total as well as differential cross 
section were obtained in~\cite{bea}. There is however a substantial 
difference between our approach and that of Ref.~\cite{bea} 
or simple decomposition in powers of $\ln(s)$, namely that 
we consider the opening channels (in $s$) as threshold effects, the 
relevant prongs being separated in rapidity by $\ln s_0$, $s_0$ being a
parameter related to the average subenergy in the ladder. Although such 
an approach inevitably introduces new parameters, we consider it more 
adequate in the framework of the finite-ladder approach. 
We mention these attempts only for the sake of completeness, although we stick
to the simplest case of $t=0$, where there are hopes to have some connection
with the QCD calculations.

In Section 2 we consider a new parametrization for total cross
sections based on the contribution from a finite series of QCD diagrams with
relative weights (coefficients) and rapidity gaps to be determined from the
data. Each set of the diagrams is "active" in "its zone", i.e. the
parameters should be fitted in each energy interval separately and the
relevant solutions should match. The matching procedure will be similar to
that known for the wave functions in quantum mechanics, i.e. we require
continuity of the total cross section and of its first derivative.
In Section 3 we present the result of fits to the $p\bar p$ and $pp$ 
experimental data. Section 4 is devoted to a discussion of the truncated 
series in QCD and to the calculation of the coefficients of the powers
of $\ln s$. Finally, in Section 5 we will draw our conclusions.

\section{Description of the model}

The Pomeron contribution to the total cross section is represented in the form
\begin{equation}
\sigma_t^{(P)}(s)=\sum_{i=0}^N f_i\:\theta(s-s_0^i)\:\theta(s_0^{i+1}-s)\;,
\label{z1}
\end{equation}
where
\begin{equation}
f_i=\sum^i_{j=0}a_{ij}L^j\;,
\label{z2}
\end{equation}
$s_0$ is the prong threshold, $\theta(x)$ is the step function and 
$L\equiv \ln(s)$. Here and in the following, by $s$ and $s_0$  
respectively, $s$/(1\,GeV$^2$) and $s_0$/(1\,GeV$^2$) is implied. 
The main assumption in Eq.~(\ref{z1}) is that the widths of the rapidity gaps 
$\ln(s_0)$ are the same along the ladder. The functions $f_i$ are polynomials 
in $L$ of degree $i$, corresponding to finite gluon ladder
diagrams in QCD, where each power of the logarithm collects all the relevant
diagrams. When $s$ increases and reaches a new threshold, a new prong opens 
adding a new power in $L$. In the energy region between two neighbouring 
thresholds, the corresponding $f_i$, given in Eq.~(\ref{z1}), is supposed 
to represent adequately the total cross section.

In Eq.~(\ref{z1}) the sum over $N$ is a finite one, since $N$ is proportional to 
$\ln(s)$, where $s$ is the present squared c.m. energy. Hence this model 
is quite different from the usual approach where, in the limit $s \to
\infty$, the infinite sum of the leading logarithmic contributions gives 
rise to an integral equation for the amplitude.

To make the idea clearer, we describe the mechanism in the case of
three gaps (two rungs).
To remedy the effect of the first threshold and get a smooth behavior at low
energies, we have included also a Pomeron daughter, going like $\sim 1/s$ 
in the first two gaps with parameters $b_0$ and $b_1$
respectively. Then
\begin{eqnarray}
f_0(s)  = a_{00}+b_0/s  && \;\;\;\mbox{for}\qquad s \leq s_0\;, \\
f_1(s)  = a_{10}+b_1/s+a_{11}L  && \;\;\;\mbox{for}\qquad s_0\leq s \leq 
s_0^2\;,\\
f_2(s)  = a_{20}+a_{21}L+a_{22}L^2  &&  \;\;\;\mbox{for}\qquad s_0^2\leq s \leq 
s_0^3\;. 
\end{eqnarray}
By imposing the requirement of continuity (of the cross section and of its
first derivative) one constrains the parameters. E.g., from the 
conditions $f_1(s_0)=f_0(s_0)$ and $f'_1(s_0)=f'_0(s_0)$ the relations
\begin{displaymath}
b_1=a_{11} s_0+b_0\;,
\end{displaymath}
\begin{displaymath}
a_{10}=a_{00}-a_{11}\ln (s_0)-a_{11}
\end{displaymath}
follow. Furthermore, from $f_2(s_0^2)=f_1(s_0^2)$ and 
$f'_2(s_0^2)=f'_1(s_0^2)$ one gets
\begin{displaymath}
a_{20}=a_{22} \ln^2(s_0^2)+a_{10}+b_1 (1+\ln(s_0^2))/s_0^2\;,
\end{displaymath}
\begin{displaymath}
a_{21}=a_{11}-2a_{22}\ln (s_0^2)-b_1/s_0^2\;.
\end{displaymath}
The same procedure can be repeated for any number of gaps. 

In fitting the model to the data, we rely mainly on $p\bar p$ data that
extend to the highest (accelerator) energies, to which the Pomeron is
particularly sensitive. To increase the confidence level, $pp$ data were
included in the fit as well. To keep the number of the free parameters as
small as possible and following the successful phenomenological approach of
Donnachie and Landshoff~\cite{DL}, a single ''effective'' Reggeon trajectory 
with intercept $\alpha \left( 0\right) $ will account for nonleading 
contributions, thus leading to the following form for the total cross 
section:
\begin{equation}
\sigma _t(s)=\sigma_t ^{(P)}(s)+R(s)\;,
\end{equation}
where $\sigma_t^{(P)}(s)$ is given by Eq.~(\ref{z1}) and $R(s)=a s^{\alpha 
(0)-1}$ (the parameter $a$ is different for $p\bar p$ and
$pp$ and is considered as an additional free parameter).

Ideally, one would let free the width of the gap $s_0$ and consequently the
number of gluon rungs (highest power of $L$). Although possible, technically
this is very difficult. Therefore we considered only the cases of two and three 
rungs and, for each of them, we treated $s_0$ as a free parameter.

Notice that the values of the parameters depend on the energy range of the
fitting procedure. For example, the values of the parameters in $f_0$ if
fitted in "its" range, i.e. for $s\leq s_0$, will get modified in $f_1$ with the
higher energy data and correspondingly higher order diagrams included.

\section{Fits to the $p\bar p$ and $pp$ data}

As a first attempt, only three rapidity gaps, that 
correspond to two gluon rungs in the ladder were considered.
Fits to the $p\bar p$ and $pp$ data were performed from $\sqrt{s}=4$ GeV up
to the highest energy Tevatron data (for $p\bar p$), including all the results
from there~\cite{abe}. The resulting fit is shown in Fig.~2 with a 
$\chi^2$/d.o.f. $\approx 1.71$. The values of the fitted parameters are quoted 
in Table~1. Interestingly, the value of $s_0$ turns out to be very close to 144 
GeV$^2$, i.e. the value for which the energy range considered is covered with 
equal rapidity gaps uniformly.

Next, we covered the energy span available in the accelerator
region by four gaps, resulting in 3 gluon rungs and consequently $L^3$ as the
maximal power. After the matching procedure, we are left with ten free parameters: 
first of all $s_0$, then $a_{00},b_0,a_{11},a_{22},a_{32}$, $a_{33}$, each 
determined in its range, while the two $a$'s and $\alpha \left( 0\right)$ are 
fitted in the whole range of the data. The final value for $s_0$ turned out to 
be $s_0\simeq 42.5$ GeV$^2$ resulting in a sequence of energy intervals ending 
at $\sqrt{s}=1800$ GeV.
It is amusing to notice that a search for the phase space region where 
the production amplitude in the multicluster configuration has a maximum 
leads, with the help of cosmic ray data, to an average "subenergy" $<s_i> 
\sim 44$ GeV$^2$~\cite{BP}, that is very near to the value of $s_0$ found 
in the fit.

Fig.~3 shows our fit to the $p\bar p$ and $pp$ total cross section data. The
values of the fitted parameters are quoted in Table~2. The value of the $\chi
^2/$d.o.f. is $\sim 1.38$, much better than in the case of two gluon 
rungs. It is interesting to observe that the coefficients in front of the leading
logarithms are related roughly by a factor of 1/10.
The value of the effective Reggeon intercept remains rather low, close to
0.45, comparable with the value found in Ref.~\cite{Kaidalov}.

\section{Explicit iterations of BFKL}

From the theoretical point of view, the phenomenological model of 
Section 2 corresponds to the explicit evaluation in QCD of gluonic 
ladders with an increasing number of $s$-channel gluons.
This correspondence is far from literal since each term of the BFKL
series takes into account only the dominant logarithm in the limit
$s\to \infty$. In the following 
we give concrete expressions for the forward high energy scattering 
amplitudes for photons and hadrons in the form of an expansion in powers of large 
logarithms in the leading logarithmic approximation.

We start from known results obtained in paper~\cite{BL} where an 
explicit expression for the total cross section for hadron-hadron scattering
has been obtained. In the high energy limit, it is convenient to 
introduce the Mellin transform of the amplitude   
\begin{displaymath}
A(\omega,t)=\int_0^\infty d\left(\frac{s}{m^2}\right)
\left(\frac{s}{m^2}\right)^{-\omega-1}\frac{\mbox{Im}_s A(s,t)}{s}
\end{displaymath}
and its inverse
\begin{displaymath}
\frac{\mbox{Im}_s A(s,t)}{s}=\frac{1}{2\pi i}
\int_{\delta-i\infty}^{\delta+i\infty} d\omega
\left(\frac{s}{m^2}\right)^\omega A(\omega,t)\;.
\end{displaymath}
The general expression of $A(\omega,t)$ in the leading logarithmic approximation has 
the form:
\begin{displaymath}
A(\omega,t)=\int d^2k \frac{\Phi^a(k,q) \: F_\omega^b(k,q)}{k^2(q-k)^2}\;,
\end{displaymath}
where $\Phi^a(k,q)$ and  $\Phi^b(k,q)$ (see next equation) are the impact factors of the colliding 
hadrons $a$ and  $b$, obeying the gauge conditions $\Phi^j(0,q)=\Phi^j(q,q)=0$ $(j=a,b)$.
The quantity $F_\omega^b(k,q)$ obeys the BFKL equation:
\begin{displaymath}
\omega F_\omega^b(k,q)=\Phi^b(k,q)+\gamma \int 
\frac{d^2k'}{2\pi} 
\frac{A(k,k',q)F_\omega^b(k',q)-B(k,k',q)F_\omega^b(k,q)}{(k-k')^2}\;,
\end{displaymath}
with
\begin{displaymath}
A(k,k',q)=\frac{-q^2(k-k')^2+k^2(q-k')^2+k^{'2}(q-k)^2}{k^{'2}(q-k')^2}\;,
\end{displaymath}
\begin{displaymath}
B(k,k',q)=\frac{k^2}{k^{'2}+(k'-k)^2}+\frac{(q-k)^2}{(q-k')^2+(k-k')^2}\;.
\end{displaymath}
and
\begin{displaymath}
\gamma=3 \frac{\alpha_s}{\pi}\;.
\end{displaymath}
The strong coupling $\alpha_s$ is assumed to be frozen at a suitable 
scale set, for example, by the external particles.
The iteration procedure and the reciprocal Mellin transform give (besides we 
put $q=0$):
\begin{displaymath}
\sigma_t(s)=\frac{\mbox{Im}_s A(s,0)}{s}=\int 
d^2k \frac{\Phi^a(k,0)}{(k^2)^2}\: \left[\Phi^b_0(k)+\rho 
\Phi_1^b(k)+\frac{1}{2!}\rho^2\Phi^b_2+...\right]\;,
\end{displaymath}
where 
\begin{equation}
\rho=\frac{3\alpha_s}{\pi}\ln\left( \frac{s}{m^2}\right)
\label{zr}
\end{equation}
and the subsequent iterations begin from $\Phi_0^b(k)=\Phi^b(k,0)$.
In the previous integral and everywhere in the following, all the momenta
are 2-dimensional Euclidean vectors, living in the plane transverse
to the one formed by the momenta of the colliding particles.

For the case of photon-photon scattering one has~\cite{BL} (see also the
references quoted there)
\begin{displaymath}
\Phi^\gamma=\sum_{i=1}^2\tau^{\gamma\gamma}(k,0)=\frac{2}{3}\alpha\alpha_s
T\left(\frac{k^2}{m^2}\right)\;,
\end{displaymath}
where
\begin{displaymath}
T\left(\frac{k^2}{m^2}\right)=\frac{1}{4}
+\frac{5-\beta^2}{8\beta}\ln\left(\frac{\beta+1}{\beta-1}\right)\;,
\end{displaymath}
\begin{displaymath}
\beta^2=1+\frac{4m^2}{k^2}>1\;.
\end{displaymath}
From the integrals\footnote{S.~Gevorkyan, private communication}
\begin{displaymath}
\frac{1}{\pi}\int d^2k \: \frac{T^2\left(\frac{k^2}{m^2}\right)}{(k^2)^2} 
= \frac{0.673}{m^2}\;,
\end{displaymath}
\begin{displaymath}
\int_0^\infty \frac{d t}{t}\int_0^\infty \frac{dt'}{t'}T(t)
\left[\frac{T(t')-T(t)}{|t-t'|}+\frac{T(t)}{\sqrt{t^2+4t^{'2}}}\right]=
\frac{1.443}{m^2}\;,
\end{displaymath}
we conclude 
\begin{equation}
\sigma^{\gamma\gamma \to 2 q 2 \bar q}(s)=\sigma_0\left[1+6.4
\frac{\alpha_s}{\pi}
\ln\left(\frac{s}{m^2}\right)\right]\;,
\label{zs}
\end{equation}
where $\sigma_0$ is a constant and we used the approximated
equality $(3\times 1.44)/0.67=6.4$.

To obtain the cross section of proton-proton scattering, we use the ansatz
of Ref.~\cite{levin} for the impact factor of a hadron 
in terms of its form factor $F(q^2)$:
\begin{displaymath}
\Phi^p(k,q)=F^p\left(\frac{q^2}{4}\right)
-F^p\left(\left(k-\frac{q}{2}\right)^2\right)\;,\;\;\; 
\Phi^p(0,q)=\Phi(q,q)=0\;.
\end{displaymath}
Here the 2-dimensional Euclidean vector $q$ is related to the 4-dimensional 
transferred momentum $Q$ by the relation $Q^2=-q^2<0$.
Using the formulae given above, we obtain, for a simplified
but experimentally acceptable form of proton's form factor,
\begin{equation}
\Phi_0(k)=a k^2 e^{-bk^2}\;,
\label{zf}
\end{equation}
where $a$ and $b$ are in GeV$^{-2}$. It is convenient to define
\begin{displaymath}
\psi_n(k^2)=\frac{\Phi_n(k)}{k^2}\;,
\end{displaymath}
then (for $n\geq 1$)
\begin{displaymath}
\psi_n(k^2)=\int_0^1\frac{dx}{1-x}\biggl(\psi_{n-1}(k^2x)-\psi_{n-1}(k^2)\biggr)
+\int_1^\infty 
\frac{dx}{x-1}\left(\psi_{n-1}(k^2x)-\frac{1}{x}\psi_{n-1}(k^2)\right)
\end{displaymath}
and
\begin{equation}
\sigma_t(s)=\pi 
\int_0^{\infty}\,dk^2\,\psi_0(k^2)\sum_n\psi_n(k^2)\frac{\rho^n}{n!}\;.
\label{zz}
\end{equation}
The integrations can be performed analytically, due to the simple 
choice of the impact factor in Eq.~(\ref{zf}), and the final result is:
\begin{displaymath}
\sigma_t(s)=\frac{\pi a^2}{2b} \left\{ 1+2 (\ln 2)\rho+\left[\frac{\pi^2}{12}+
2 (\ln 2)^2 \right]\rho^2+ \right.
\end{displaymath}
\begin{equation}
\left. \frac{1}{3} \left[\frac{\pi^2}{2}(\ln 2)+4 (\ln 
2)^3-\frac{3}{4}\zeta(3)\right] \rho^3+\ldots \right\}\;.
\label{za}
\end{equation}
where $\rho$ is defined in Eq.~(\ref{zr}).

As stressed above, the coefficients of different powers of 
$\ln(s/m^2)$ in Eq.~(\ref{za}) refer to the dominant contribution, at asymptotic
energies, for each perturbative order. In the fit, instead, the 
Pomeron contribution is determined only from the experimental data
at high but finite energies. However, we can obtain 
a rough estimate of the importance of the subleading 
contributions by comparing Eq.~(\ref{za}) with the phenomenological fit 
of the previous Section, in particular with the amplitude $f_3$ relative 
to the last gap of the three rungs case. If we assume  a commonly used 
value for the strong coupling, $\alpha_s \sim 0.5-0.7$~\cite{levin}, we
must conclude that subleading contributions to the QCD Pomeron are
important. This finding may be related to the fact that energies reached 
by the present accelerators are not yet asymptotic.

\section{Conclusions}

Although high quality fits were not the primary goal of the present study,
we still may conclude that they are compatible with those 
existing~\cite{DL}, and there is still room for further improvement.
Our main goal instead was to seek for an adequate picture of the Pomeron
exchange at $t=0$.  In our opinion, it is neither an  infinite sum of 
gluon ladders as in the BFKL approach~\cite{FKL,BL,L}, nor its power 
expansion. In fact, the finite series - call it "threshold approach" - 
considered in this and our previous paper~\cite{fiore} realizes a non 
trivial dynamical balance between the total reaction energy and the 
subenergies equally partitioned between the multiperipheral ladders.

An important finding of the present paper, confirmed both by the fits to 
the $pp$ and $p\bar p$ cross sections (Sec. 3) and by the QCD 
calculations (Sec. 4) is the rapid convergence of the series.

''Footprints'' of the prongs at low energies are slightly visible in Fig.~2
(especially in the case of $pp$ scattering where the contribution 
from secondary Reggeons is smaller than in $p\bar p$). A more detailed study of this
phenomenon may answer the question whether this is merely an artifact or a
manifestation of Pomeron's basic properties.
The quality of the present fits will be definitely improved when future 
data from RHIC and LHC will be available.
As to the QCD calculations, their precision and efficiency are biased at 
finite energy by difficulties in estimating non-asymptotic terms, 
neglected here and elsewhere. As a consequence, e.g. in the diagram giving 
a $\ln^2(s)$ contribution, the constant and $\ln(s)$ terms (see 
Fig.~(1)) are neglected. 

In any case, few terms of powers of $\ln(s)$ are sufficient to describe 
the data at all realistic energies. As a guess, we do not exclude that 
higher terms cancel completely, but anyway they are negligibly small. The 
case of two terms (logarithmic rise in $s$) is particularly interesting 
as it corresponds to a dipole Pomeron with a number of attractive 
features~\cite{JShS} such as self-reproducibility with respect of 
unitarity corrections. In case of a $\ln^2(s)$ rise (three terms) we 
still should not worry about the Froissart bound, so ultimately the 
Pomeron as viewed in this paper does not need to be unitarized. This 
conclusion is an important by-product of our paper. For the dipole 
Pomeron, relevant calculations for $t\neq 0$ are interesting and important but 
difficult. In the case of a single gluon rung they were performed in 
Ref.~\cite{bea} and, with a non-perturbative gluon propagator, in the
last reference of~\cite{Francesco}. 

The role and the value of the width of the gap, $s_0$, is an important 
physical parameter {\sl per se}, independent of the model presented 
above. We have fitted it and compared successfully with the prediction 
from cosmic-ray data. However its value may be estimated e.g. as the 
lowest energy where the Pomeron exchange is manifest, although the 
latter is also a matter of debate.
Further fits of the model to new experimental data may settle some details left open
by this paper.

\section{Acknowledgment}

We thank V. Fadin, A. Kaidalov and L. Lipatov for numerous discussions on the
Pomeron. One of us (L.J.) is grateful to the Dipartimento di Fisica 
dell'Universit\`a della Calabria and to the Istituto Nazionale di Fisica 
Nucleare - Sezione di Padova e Gruppo Collegato di Cosenza for their 
warm hospitality and financial support. The work of L.J. was partly 
supported by INTAS, grant 97-1696 and CRDF, grant UP1-2119.

\newpage

\centerline{\bf TABLES}

\vspace{3cm}

\begin{table}[h]
\centering
\begin{tabular}{||c|c|c|c||}
\hline
$s_0$ & $\alpha \left( 0\right)$ & $a_{p\bar p}^{}$ & $a_{pp}^{}$ \\
\hline
$147.97(93)$ & $0.441(11)$ & $71.7(3.4)$ & $0.00(37)$ \\
\hline
\hline
$b_0$ & $a_{00}$ & $a_{11}$ & $a_{22}$ \\
\hline
$35.6(1.5)$ & $38.097(23)$ & $2.300(38)$ & $0.857(61)$ \\
\hline
\hline
$a_{10}$ & $a_{20}$ & $a_{21}$ & $\chi^2/$d.o.f. \\
\hline
24.3 & 110.1 & -14.85 & 1.71 \\
\hline
\end{tabular}
\caption{Value of the parameters in the case of 2 rungs.  The parameters 
$b_i$ and $a_{\ldots}$ are given in units of 1 mb. The quantities
in round parenthesis represent the errors. Parameters without error are 
derived from the matching condition.}
\end{table}

\vspace{2cm}

\begin{table}[h]
\centering
\begin{tabular}{||c|c|c|c|c|c||}
\hline
$s_0$ & $\alpha(0)$ & $a_{p \bar p}^{}$ & $a_{pp}^{}$ & $b_0$ & $a_{00}$ \\
\hline
$42.43(53)$ & $0.4295(95)$ & $160.9(8.6)$ & $85.2(6.6)$ & $-180(18)$ & $33.20(29)$ \\
\hline
\hline
$a_{11}$ & $a_{22}$ & $a_{32}$ & $a_{33}$ & $a_{10}$ & $a_{20}$ \\
\hline
$2.631(86)$ & $0.324(55)$ & $0.19(22)$ & $0.000(38)$ & 20.7 & 38.6 \\
\hline
\hline
$a_{21}$ & $a_{30}$ & $a_{31}$ & $b_1$ & $\chi^2/$d.o.f. & \\
\hline
-2.19 & 22.3 & 0.72 & -68.42 & 1.38 & \\
\hline
\end{tabular}
\caption{Value of the parameters in the case of 3 rungs. The parameters 
$b_i$ and $a_{\ldots}$ are given in units of 1 mb. The quantities
in round parenthesis represent the errors. Parameters without error are 
derived from the matching condition.}
\end{table}

\newpage

\phantom{.}

\begin{figure}[htb]

\begin{picture}(400,200)(0,0)

\CCirc(50,150){15}{Black}{Black}
\Line(25,175)(75,125)
\Line(25,177)(75,127)
\Line(25,125)(75,175)
\Line(25,123)(75,173)

\Line(125,175)(175,175)
\Line(125,177)(175,177)
\ZigZag(140,125)(140,175){3}{5}
\ZigZag(160,175)(160,125){3}{5}
\Line(125,125)(175,125)
\Line(125,123)(175,123)
\Vertex(140,124){2}
\Vertex(140,176){2}
\Vertex(160,124){2}
\Vertex(160,176){2}

\Line(225,175)(275,175)
\Line(225,177)(275,177)
\ZigZag(240,125)(240,175){3}{5}
\ZigZag(260,175)(260,125){3}{5}
\Line(225,125)(275,125)
\Line(225,123)(275,123)
\Vertex(240,124){2}
\Vertex(240,176){2}
\Vertex(260,124){2}
\Vertex(260,176){2}
\Vertex(240,150){4}
\Vertex(260,150){4}
\Gluon(240,150)(260,150){3}{2}

\Text(100,150)[c]{{\Large =}}
\Text(200,150)[c]{{\Large +}}
\Text(300,150)[l]{{\Large + $\;\;\;\;\; \cdots$}}

\Line(25,75)(75,75)
\Line(25,25)(75,25)
\Vertex(50,75){2}
\Vertex(50,25){2}
\Gluon(50,25)(50,75){3}{5}
\Vertex(50,50){4}
\Gluon(50,50)(75,50){3}{3}

\Line(125,75)(175,75)
\Line(125,25)(175,25)
\Vertex(150,75){2}
\Vertex(150,25){2}
\Gluon(150,25)(150,75){3}{5}
\Vertex(150,50){2}
\Gluon(150,50)(175,50){3}{3}

\Line(225,75)(275,75)
\Line(225,25)(275,25)
\Vertex(250,75){2}
\Vertex(250,25){2}
\Gluon(250,25)(250,75){3}{5}
\Vertex(235,75){2}
\Gluon(235,75)(266,60){3}{5}

\Text(100,50)[c]{{\Large =}}
\Text(200,50)[c]{{\Large +}}
\Text(300,50)[l]{{\Large + $\;\;\;\;\; \cdots$}}

\end{picture}
\caption{Schematic representation of the total cross section in the 
leading $\ln(s)$ approximation (first row). Double lines represent 
protons or anti-protons, vertical zig-zag lines are Reggeized
gluons, horizontal wavy lines are gluons. The effective vertex for
two Reggeized gluons and one gluon is defined in the second row. Here external
lines here can represent quark or gluons.}
\end{figure}
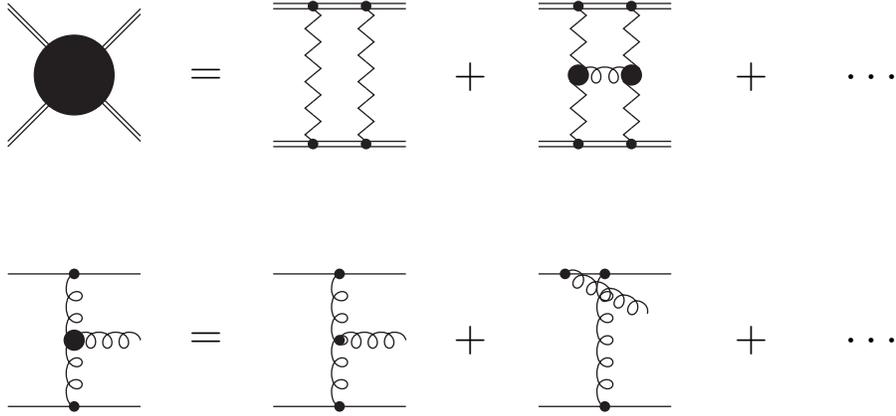

\begin{figure}[htb]
\begin{center}
{\parbox[t]{5cm}{\epsfysize 12cm \epsffile{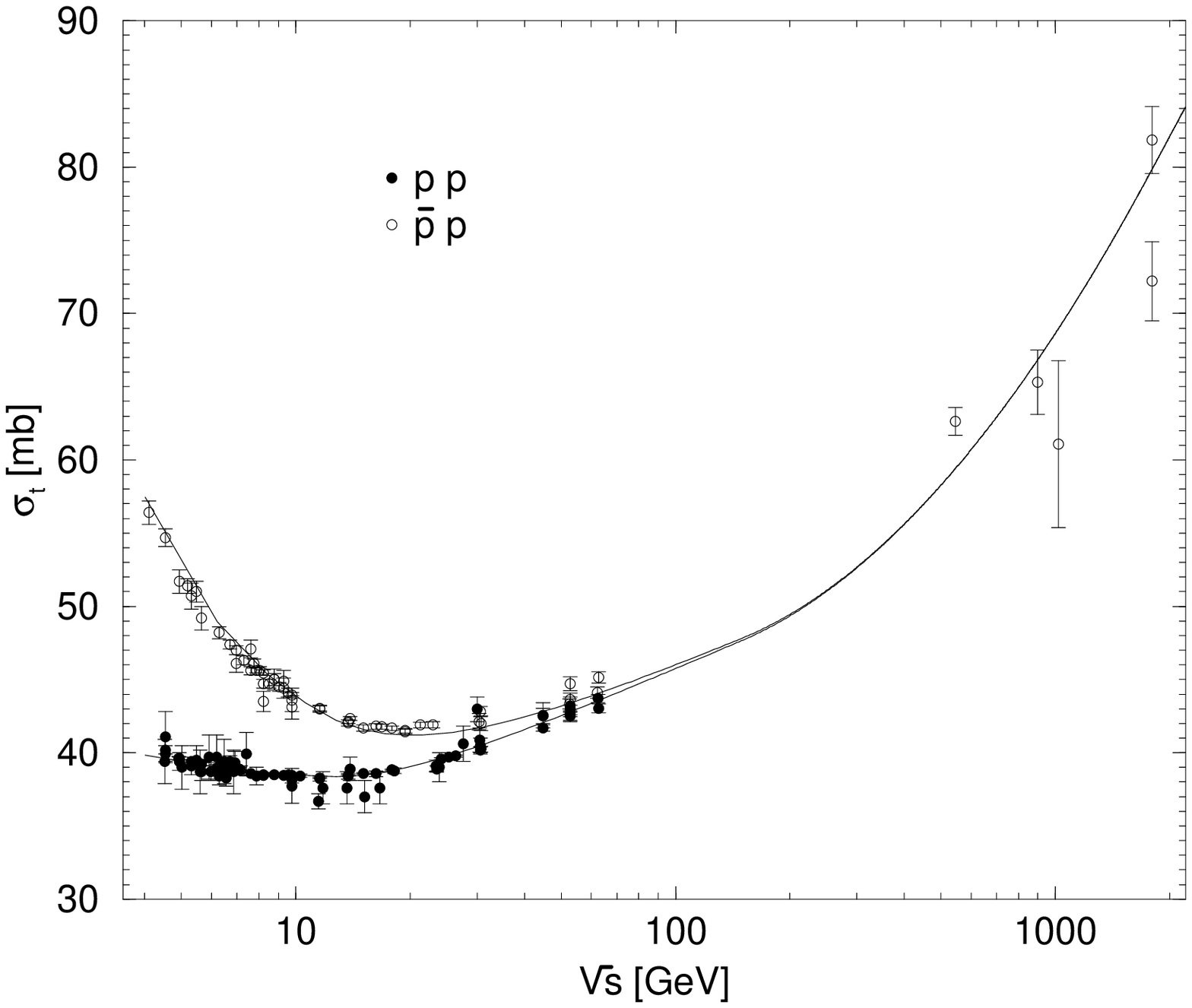}}}
\end{center}
\caption[]{Total cross section calculated up to 2 gluon rungs and fitted to
the $p\bar p$ and $pp$ data.}
\end{figure}

\begin{figure}[htb]
\begin{center}
{\parbox[t]{5cm}{\epsfysize 12cm \epsffile{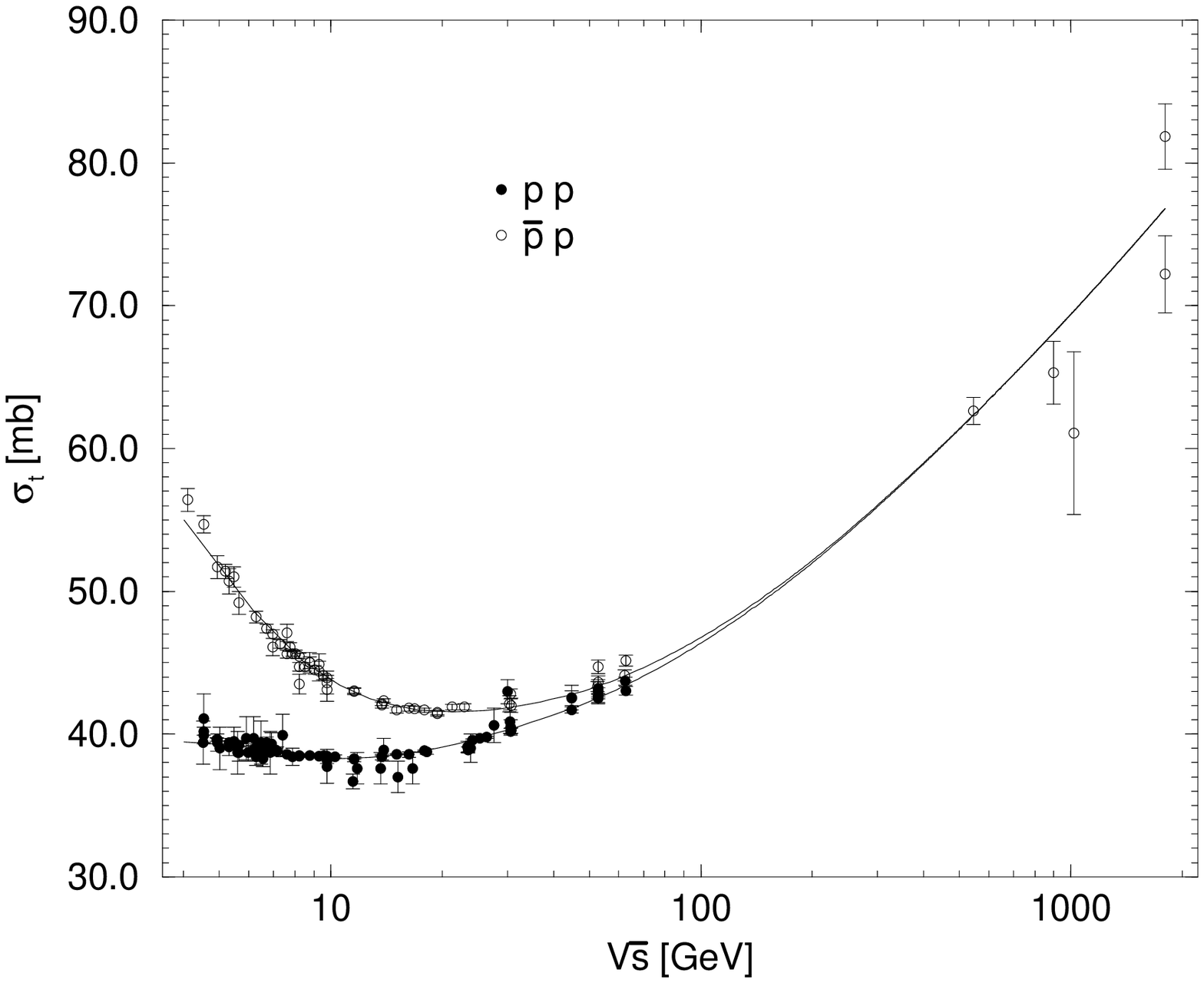} }}
\end{center}
\caption[]{Total cross section calculated up to 3 gluon rungs and fitted to
the $p\bar p$ and $pp$ data.}
\end{figure}

\newpage

\end{document}